\documentclass{PoS}

\usepackage{amssymb}
\usepackage{fontenc}
\usepackage{times}
\usepackage{mathptmx}
\usepackage{graphicx}

\def\a{\alpha}
\def\b{\beta}
\def\g{\gamma}

\def\m{\mu}

\def\l{\lambda}

\newcommand{\be}{\begin{equation}}
\newcommand{\ee}{\end{equation}}
\newcommand{\bea}{\begin{eqnarray}}
\newcommand{\eea}{\end{eqnarray}}
\newcommand{\bal}{\begin{aligned}}
\newcommand{\eal}{\end{aligned}}

\newcommand{\eq}[1]{Eq.~(\ref{#1})}

\newcommand{\tr}{\mathrm{tr}\,}

\usepackage{tikz}
\usetikzlibrary{positioning,arrows}
\usetikzlibrary{decorations.pathmorphing}
\usetikzlibrary{decorations.markings}

\title{Non-perturbative Gauge-Higgs Unification, a quantum and bosonic mechanism of spontaneous gauge symmetry breaking}

\ShortTitle{Non-perturbative Gauge-Higgs Unification}

\author{\speaker{Nikos Irges}
       \thanks{Dedicated to Yannis Bakas.}\\
      Department of Physics\\
      National Technical University of Athens\\
      Zografou Campus, GR-15780 Athens, Greece\\
      E-mail: \email{irges@mail.ntua.gr}}

\abstract{We review a construction called Non-Perturbative Gauge-Higgs Unification which is a
rather rare type of Higgs mechanism that is a pure quantum and bosonic effect.
It is constructed on a five-dimensional "orbifold lattice" with $SU(2)$ bulk gauge symmetry
and a (spontaneously broken) Abelian-Higgs like theory on the boundaries.}

\FullConference{Corfu Summer Institute 2016 "School and Workshops on Elementary Particle Physics and Gravity"\\
		31 August - 23 September, 2016\\
		Corfu, Greece}

\begin{document}

\section{Introduction}

Particle physics models are Quantum Field systems defined by the following data:
Field content, symmetries, dimensionality of space-time and some signs.
Physical observables are then computed (in Euclidean space-time) from a path integral, according to
\be\label{O}
\langle {\cal O} \rangle \sim \int D\Phi \, {\cal O} \, e^{-S[\Phi;g]}
\ee
where $\Phi$ is a generic set of fields, $S[\Phi;g]$ is a gauge invariant action consisting of a Lagrangean
of engineering dimension d integrated over $d$-dimensions.
The Lagrangean is constructed from the fields $\Phi$ and couplings $g$ following the "anything not prohibited appears" rule
and $\cal O$ is a gauge invariant operator. Here we will restrict our attention to $d\ge 4$.

Quantum Field Theory provides the rules according to which \eq{O} can be computed.
We will be thinking of such computations being done with an ultraviolet (UV) cut-off.
There is one more piece of data that may be missing before proceeding with the actual computation.
It is the signs in front of the interaction terms among the various fields which are not always fixed by
the symmetries. This is the case for example in the Higgs sector of the Standard Model (SM) where in the potential
\be
V(\Phi) = -\m^2 \Phi^\dagger\Phi + \l \left( \Phi^\dagger\Phi \right)^2\, ,
\ee
(with $\m^2>0, \l>0$), the relative sign between the two terms is arbitrary.
Apart from this sign, its form is completely fixed as long as we 
do not parametrise any new physics or quantum corrections by including higher engineering dimension operators.
Notice that if the relative sign was positive, the vacuum expectation value of the Higgs would be zero
and there would be no Higgs mechanism. 
Experiment however has shown us that we live in a state described by a relative minus sign.
Also by experiment, we know that the couplings in the Electroweak sector are weak
and this allows us to compute everything in the context of (weak coupling) perturbation theory.
A standard assumption is that all non-perturbative effects in the SM originate from the strong interactions 
and those can be included in a systematic way in physical processes.

The only place in the SM where this conceptually simple algorithm breaks down is associated with the quantum corrections to the
Higgs mass that (perturbatively) develops a quadratic cut-off dependence. 
This is believed to generate a hierarchy problem when the cut-off is allowed to take large values.
The situation is further blurred if we recall that in the absence of fermions the coupling $\l$ increases monotonically towards the
UV and (much before it hits a Landau pole) can turn non-perturbative or even hit a phase transition while one is still grappling with fine tuning issues.\footnote{
The presence of fermions adds one more complication, that of the quantum instability originating from 
the top Yukawa coupling that tends to decrease $\l$ and eventually turn it negative.
This is a nuisance that we will neglect in this discussion.}
The "normal" hierarchy of scales from the IR to the UV that we implied, i.e. scale where fine tuning becomes unnatural, scale where the coupling turns non-perturbative,
scale where a phase transition occurs and the Landau pole scale,
could be disrupted in extensions of the SM.

Before we continue with the description of our extra-dimensional model, let us summarise various ways of
generating an effective Higgs mechanism.
These can be divided in two large classes. One is the class where
there is a classical potential inserted in the action that triggers the breaking of some of the gauge symmetry.
The other class is where the presence of fermions is necessary to trigger a Higgs mechanism.
Examples of models that belong to the first class are the ones that generate a Higgs potential via
the compactification of more than one extra dimensions, through the term $F_{ij}^2$, with
$i,j$ labelling the components of the gauge field strength tensor in the extra dimensions. 
Viewed from four dimensions, if the system is dimensionally reduced, this is an effective, gauge invariant, 
classical potential for the Higgs-like scalars $A_i$, the gauge potential fields along the extra dimensions. 
A second type of models with classical Higgs potentials consists of four-dimensional extended Higgs sector models, such as the MSSM and variations.
The example in this class that is closer in spirit however to the one discussed here is the Coleman-Weinberg (CW) mechanism \cite{CW}
which is an interplay between a classical quartic and a quantum quadratic term.
The Higgs potential is the sum of these two terms thus it is a semi-quantum effect from our point of view.

Even though we said that we will not consider dimensions lower than four, it is hard to resist
mentioning as a primary example of the second class of models, superconductivity.
In superconductivity the appearance of an effective scalar field is a weak coupling but non-perturbative effect due to the 
condensation of Cooper electron pairs. The phenomenon can be then described by a bosonic effective action
where the scalar potential appears as a classical object, as in the SM.
In any case, the mechanism at the microscopic level is tied to fermions.
Regarding higher dimensional examples of this fermionic class a popular one
is the Hosotani mechanism \cite{Hosotani} in compactified higher dimensional gauge theories \cite{Kubo}. 
According to this mechanism there is a quantum, CW type potential generated for the scalars $A_i$.
An interesting special subclass is $d=5$ compactified gauge theories, where since the
potential vanishes at the classical level due to the absence of a $F_{55}^2$ term, the scalar
potential is a purely quantum object. In fact it is relatively simple to compute it for a $d=5$ $SU(N)$ gauge theory
compactified on a circle (or an interval; the result is similar) where one can show that the resulting potential has the form \cite{Ignatios}
\be\label{Hosotani}
V \sim \sum_{n=1}^\infty \frac{\cos{(2\pi n \a)}}{n^5}\, 
\ee
with $\a$ parametrising a vacuum expectation value. The minimization of this potential in the absence of fermions
shows that $\a$ is an integer and therefore that the Kaluza-Klein tower associated with the
gauge boson field with masses $m_n = \frac{n+\a}{R}$ (with $R$ the length of the fifth dimension), 
undergoes spectral flow, leaving the original
$SU(N)$ gauge symmetry intact. If however one adds enough number of fermions, the vev $\a$ can take
non-integer values thus breaking the gauge symmetry into a subgroup of $SU(N)$.
Another example in this fermionic class of models that we would like to mention is the one where
the Higgs is a composite state of some strongly interacting non-SM fermions.
Finally, there is a number of less generic and typically more complicated models 
which rely simultaneously on classical potentials and on fermions that we do not mention.
All these are of course legitimate possibilities and only experiment can tell which (if any) will survive and which not.

To summarize, models reproducing the 4d Higgs mechanism are either classical or fermionic, or both.
The only combination that (as far as we know) did not come up until recently is "quantum and bosonic"
and this is what we would like to describe in the following.

\section{Non-perturbative Gauge-Higgs Unification (NPGHU)}

Our starting point is the example that we gave to demonstrate the Hosotani mechanism, say a pure $SU(2)$ gauge theory in $d=5$ dimensions.
Four of the dimensions can be considered to be practically infinite while the fifth dimension 
is compactified, to begin, on a circle of radius $R$. The radius is assumed to be small enough so that
the system is reduced dimensionally to four dimensions.
As mentioned, the perturbative, 1-loop CW potential for "the Higgs", $A_5$, has the form of \eq{Hosotani} and does not trigger
Spontaneous Symmetry Breaking (SSB). On the other hand, it gives a non-zero value to the Higgs mass
\be\label{mariano}
m_H^2 \sim \frac{g_5^2}{R^3}
\ee
with $g_5$ the five-dimensional $SU(2)$ gauge coupling whose square has dimension of length.
Arguably surprisingly it is finite. There is a number of non-trivial puzzles arising in this model.
From the perturbative point of view, an issue is that this model is non-renormalizable
rendering the computation questionable from the beginning.
However, non-renormalizability is supposed to result in wild infinities that we do not see.
Let us look at this in some more detail. The gauge boson mass remains zero after a trivial spectral flow and this can be 
easily attributed to the higher dimensional gauge invariance. More subtle is the Higgs potential and mass.
To begin, the typical CW computation of the potential is performed in some fixed gauge and this is why it is so simple.
In a general gauge we would observe gauge dependence, but this is already a headache in four dimensions so we do not discuss 
it any further (mass corrections though are expected to be gauge-independent). 
Second, in order to obtain the simple closed form of \eq{Hosotani} it is assumed that the Kaluza-Klein tower is infinite
because a crucial Poisson resummation is performed at some point during the computation. This is consistent only with an infinite cut-off.
Regarding the mass correction \eq{mariano}, an explicit 1-loop Feynman diagram computation \cite{Mariano} confirms finiteness and gauge-independence.
The Poisson resummation is again necessary and the computation is done in dimensional regularization with an implicit
infinite cut-off. 
There is one obvious generalization that one can try to implement, that of a finite cut-off. 
Things get immediately more obscure (and more complicated) as it is not easy to define a strictly gauge invariant
theory with a finite cut-off in the continuum. Nevertheless one can attempt to brute force a gauge symmetry breaking cut-off 
and argue that for a high enough cut-off the crime is small.
But then the KK tower can not be infinite. It must be truncated at most at the cut-off that prohibits the Poisson resummation and 
consequently a nice closed form. By doing so, the CW potential (which is again gauge dependent) still does not break
the symmetry in the pure gauge theory. If on the other hand one wants to retain full gauge invariance during the 
1-loop Higgs mass computation, it must be performed 
in some consistent finite cut-off regularization scheme such as the lattice. 
This can be done and the result \eq{mariano} is again confirmed \cite{DelDebbio}.
Finally, one can give certain believable symmetry arguments \cite{Ignatios, Mariano2} to justify the finiteness of the Higgs mass
that could possibly apply non-perturbatively.

All this certainly justifies a more systematic non-perturbative study.
We will not give a technical report of such a study, as this was done in great detail in the recent review \cite{FrancescoReview}.
We will instead focus here on some basic results, discuss certain conceptual issues and outline possible future directions of research.

\subsection{The circle, non-perturbatively}

We would like to perform a non-perturbative study of higher dimensional gauge systems with an UV cut-off in a gauge invariant way.
The lattice regularization is well suited for such a purpose. We therefore consider a 5d Euclidean hypercubic lattice, with $SU(2)$ gauge symmetry for simplicity.
The lattice is periodic in all directions, even though the radius of the usual four dimensions we will consider to be practically infinite.
The fifth dimension is a finite circle $S^1$ of radius $R$ and it gives the name to this version of the model.
The action is the standard Wilson plaquette action 

\bea
S_{\rm circle}[U] &=& \frac{\b}{2} \sum_{\rm 5d-plaq.}\, 
{\rm Re}\,\tr\,\left[1-U(p)\right] \nonumber\\
\label{circleaction}
\eea
where $\b=4a/g_5^2$ and $U(p)$ the gauge invariant elementary plaquette in the lattice whose each side has length $a$,
the so called lattice spacing. As the model stands, the two dimensionless parameters $\b$ and $N_5=2 \pi R/a$ parametrize its phase diagram.
This version of the model is not our main target. We will use it only to point out a few important properties
that we will be able to use later. 

The first important property is that the five-dimensional model, for large enough $N_5$, possesses a first order
phase transition at a value of approximately $\b_c=1.64$ \cite{Creutz}. For general $d>4$, we actually find $\b_c\simeq 6.704840/(d-1)$ \cite{George}.
The two phases that the phase transition divides are
the Coulomb phase (at $\b > \b_c$) and the Confined phase (at $\b < \b_c$).
The interesting fact about this phase transition 
is that it is not of the finite temperature type, which means that it is present even when $N_5$ is large. 
Such phase transitions are called "bulk" or sometimes "quantum" phase transitions.
Now let us see what happens as we decrease $N_5$.
There is a certain gauge invariant operator, the trace of the Polyakov loop
\be
P = {\rm tr} \prod_{{\rm links}\in\, S^1}{U}(n_5)\label{line}
\ee
constructed out of the product of all gauge links along the $5$'th direction
that can be used as an order parameter for phase transitions of the finite temperature type.
In the Coulomb phase $\langle P \rangle \ne 0 $ while in the confined phase $\langle P \rangle = 0$.
Below a critical length of the fifth dimension (and of the Polyakov loop that wraps around it), the
bulk phase transition disappears and we see a confined phase, consistently
with what we would expect from a dimensionally reduced system to four dimensions:
a 4d $SU(2)$ pure gauge theory in infinite volume has only a confined phase.
The important fact we would like to keep from here is that the non-zero vev of $P$ breaks
the global symmetry of the action, defined by the transformation
\be\label{center}
Z: \hskip .5cm U \longrightarrow z U
\ee
of links $U$ pointing in the 5th dimension, located at a fixed slice orthogonal to the 5th dimension, that multiplies them
by a center element $z\in SU(N)$ (that is just a minus sign for $SU(2)$).
But finite temperature phase transitions do not break gauge symmetries and
the continuum version of the process is just Kaluza-Klein reduction.
This is therefore the non-perturbative version of the statement that in the pure gauge theory
with periodic boundary conditions along the extra dimension there can be no Higgs mechanism.

The second important property is that the plaquette action can be generalized in a way
that (the discrete version of) Lorentz invariance is broken along the fifth dimension while it is
intact in the four-dimensional sense. This is not prohibited as long as the breaking of Lorentz invariance is not communicated 
to the whole system. That this is not the case is non-trivial in general, but there is a natural way to
do it on the lattice, via the modified, "anisotropic" action
\bea
S[U] &=& \frac{\b_4}{2} \sum_{\rm 4d-plaq.}
{\rm Re}\,\tr\,\left[1-U_4(p)\right] \nonumber\\
&+& \frac{\b_5}{2} \sum_{\rm 5d-plaq.}
{\rm Re}\,\tr\,\left[1-U_5(p)\right] \, , \label{action}
\eea
where now plaquettes on four-dimensional slices $U_4(p)$ consist of links on a lattice of lattice spacing $a_4$,
plaquettes along the extra dimension $U_5(p)$ have two of their links along the extra dimension with lattice spacing $a_5$.
The plaquettes are weighted with dimensionless gauge couplings $\b_4=4a_5/g_5^2$ and $\b_5=4a_4^2/g_5^2a_5$ respectively.
The parameters have increased from two to three: the dimensionful parameters in our lattice are now $g_5, R, a_4, a_5$ so
we can form the three dimensionless classical couplings $\b_4$, $\b_5$ and $N_5$.
The qualitatively new fact is that now there appears a regime in the phase diagram, roughly near the bulk phase transition and when $\beta_4>\beta_5$, 
where the system reduces dimensionally, not because it is compactified on a small circle but because 
four dimensional slices start fluctuating independently. 
This type of dimensional reduction is usually called dimensional reduction by localization and it has very different properties from dimensional reduction
by compactification (and presumably very different phenomenology!). 
For low enough $\b_5$, below the phase transition, the phase is called the "layered phase" \cite{Holger}. 
Keep in mind that the line of bulk phase transitions lies in a non-perturbative regime from the 5d point of view.

The third property that will also be relevant for later is that the bulk phase transition
at a given and large enough $N_5$ is of first order.
At some fixed $N_5$ the phase diagram is two dimensional and the phase transitions
fall on a line. That they are of first order can be checked by measuring via Monte Carlo simulations the expectation
value of plaquettes and from a standard cold and hot start set of measurements, observe the latent heat.
For us, the important piece of information is that any effective theory that describes the system in the vicinity
of a first order phase transition must be one with a finite cut-off that is, no continuum limit can be taken.
If this effective theory is a theory defined on a continuous space-time, its cut-off must be proportional to the inverse lattice spacing.
The explicit construction of such an effective action is (very) hard in general.
The other important, empirical fact is that masses in units of the lattice spacing
decrease as the phase transition is approached. This is consistent with the observed dimensional reduction
because the low lying spectrum near the phase transition starts to appear more and more four-dimensional.

There are rigorous lattice Monte Carlo studies of this model \cite{FrancescoReview} confirming the above three general properties
and the interested reader can find the details there.

\subsection{The orbifold, non-perturbatively}

\begin{figure*}[!ht]
\centerline{\includegraphics[width=100mm]{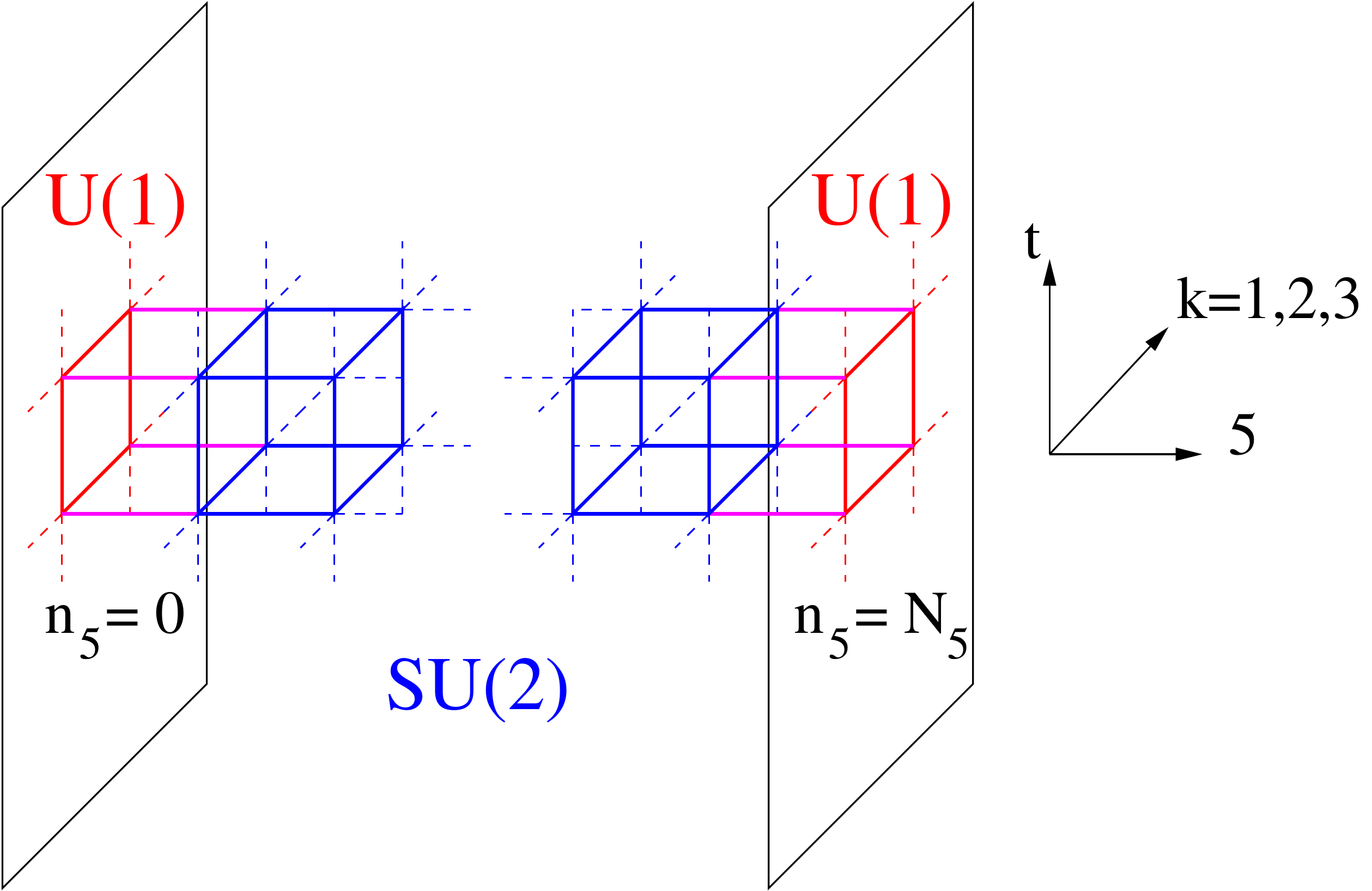}}
\caption{\small{The lattice orbifold.
\vspace{0cm}}}\label{latorb}
\end{figure*}

Because of the absence of a Higgs phase in the circle model (the deconfined phase appears to be all Coulomb) that can be traced to the
center symmetry $Z$ as the only available non-trivial relevant global symmetry, we have to modify the model.
Another, phenomenological reason to modify it is that as it is, the Higgs scalars fall in the adjoint representation
of $SU(2)$, that is a triplet. We would like to have instead a Higgs in the fundamental representation.
We can resolve simultaneously both problems by a simple trick. We project the circle in the extra dimension by the $Z_2$ transformation that identifies the upper
semicircle with the lower semicircle, turning the circle into a finite interval. Like this, we have the topology of
a finite slab whose boundaries are (infinite) four dimensional planes. In addition, we can embed this
$Z_2$ transformation in the $SU(2)$ Lie algebra in such a way that on the two boundaries only a $U(1)$
symmetry survives and out of the three scalars only two survive. 
The combination of the geometric and Lie algebraic projections produces what we call from now on the "orbifold lattice".
A picture of this construction \cite{orblat} can be seen in Fig. \ref{latorb}.
The two scalars that survive the projections on the boundaries combine in a complex scalar $H$,
that we call the Higgs. The boundary spectrum is then just the field content of the Abelian Higgs model, so to zero'th order approximation
we have an $SU(2)$ gauge symmetry in the bulk and an Abelian-Higgs model on the boundaries.
In fact, since there is an accidental reflection symmetry about the center of the interval, we can forget about one of the boundaries
and concentrate on, say, the left one.

\begin{figure*}[!ht]
\centerline{\includegraphics[width=100mm]{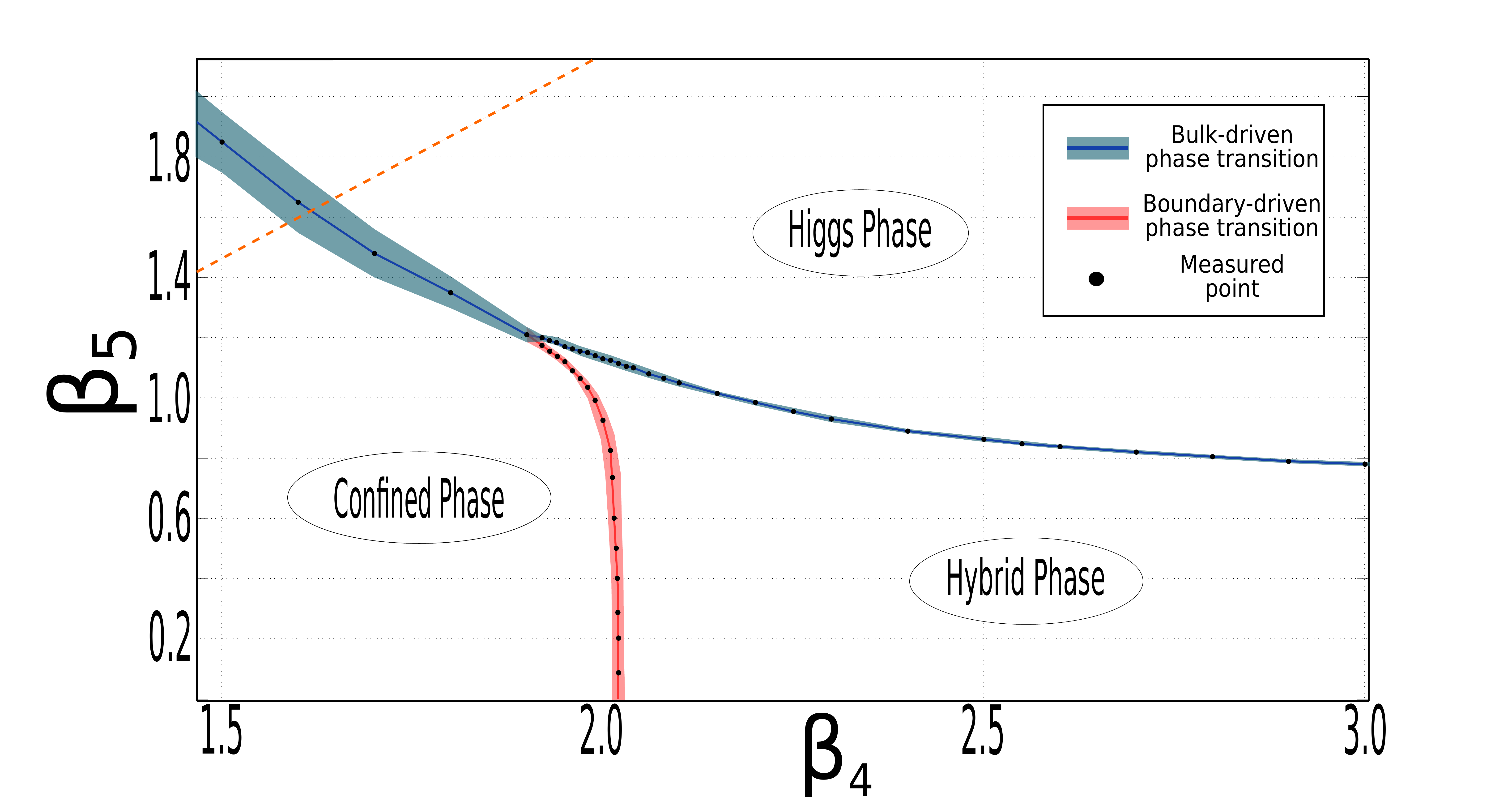}}
\caption{\small{The phase diagram of the 5d lattice orbifold model for $N_5=4$ (the picture is pretty much $N_5$-independent). 
The dotted line near the upper left corner indicates the isotropic system.
\vspace{0cm}}}\label{phase_diagram_MC}
\end{figure*}

A first observation is that the parameter space has not increased by the orbifold projection. It is still parametrized by $\b_4$, $\b_5$ and $N_5$. 
The phase diagram for the orbifold lattice can be seen in Fig \ref{phase_diagram_MC} and it was determined by Monte Carlo simulations in \cite{MaurizioPaper}
from which the figure is borrowed.

A second observation is that the second and third properties we pointed out for the circle model
pertain here exactly as before. To remind, as the Higgs-Hybrid phase transition at $\b_4>\b_5$ is approached, first the boundary 
and close enough to the phase transition \cite{MaurizioProc} also the bulk Wilson Loops oriented along the four dimensional planes, yield a static potential that
can be fitted only by a 4d Yukawa potential. This implies that the lattice decomposes into weakly interacting
four dimensional slices, or to express it in our earlier jargon, it reduces dimensionally by localization.
The first property on the other hand changes drastically. From the symmetry point of view it has to do with the disappearance of the center
symmetry transformation \eq{center} and its substitution by a new global $Z_2$ symmetry, called the "stick" symmetry \cite{stick}.
Explicitly, it acts as 
\bea
&& U(n_5=0,5) \longrightarrow g_s^{-1} U(n_5=0,5) \nonumber\\
&& U(n_5=0,\m) \longrightarrow g_s^{-1} U(n_5=0,\m) g_s
\eea
by an $SU(2)$ matrix $g_s$ on links at the "left" boundary located at $n_5=0$ and pointing along the fifth dimension and the
four dimensions respectively. This is obviously not an $SU(2)$ gauge transformation and if the matrix that implements the
orbifold projection that turns the circle theory into the interval theory is $g$, then $g_s$ is such that $\{g,g_s\}=0$.
This global symmetry has the crucial effect that the fifth dimension, no matter how small or how large,
can not feel a temperature. The bulk phase transition does not disappear ever and a certain
class of modified Polyakov loops, covariant on the interval and odd under the stick symmetry
become the order parameters for its breaking \cite{symmetries}. 
A representative of this class is the $CP$-even operator\footnote{There are also $CP$-odd representatives on which we will comment later.}\cite{orblat,MaurizioPaper}
\be\label{Zop}
{\cal Z}_k(n_0) = \frac{1}{L^3} \sum_{n_i, i=1,2,3} {\rm tr} \left[g\, U(n_\m,k) \a(n_\m+a_4{\hat k}) U^\dagger (n_\m,k) \a(n_\m)\right]
\ee
where $L$ is the four-dimensional extent of the lattice, $n_0$ is the discrete time coordinate of the operator, $k$ labels the usual spatial directions, 
$U(n_\m,k)$ are links located at $n_\m$ and pointing in the $k$-direction and $\a(n_\m)$ is
the $SU(2)$ projection of 
\be\label{Higgs}
h(n_\m) = [v(n_\m),g]
\ee
with
\be
v(n_\m) = \frac{1}{4N_5} \left(p(n_\m)-p^\dagger(n_\m)\right), \hskip 1cm p(n_\m) = l(n_\m) g l^\dagger(n_\m) g^\dagger
\ee
where $l(n_\m)$ is the from boundary to boundary line constructed from a product of links like $P$ on the circle.
It is a quite complicated object that has the quantum numbers of a four-dimensional $U(1)$ gauge boson.
Hence, when it takes a non-zero expectation value it does not only break the stick symmetry but also
generates a mass for the gauge boson. In other words, the ${\cal Z}$ operator is an order parameter for the Higgs mechanism
and as a result the deconfined phase on the interval is not a Coulomb but a Higgs phase.
This is all nicely consistent with Elitzur's theorem.
A highly non-trivial consistency check is that for given values of the dimensionless parameters $\b_4$, $\b_5$ and $N_5$, 
the measured $Z$-boson mass $a_4 m_Z$ from \eq{Zop} coincides with the
mass obtained from the fit to the 4d Yukawa potential.
It is important to also mention that the Higgs mass $a_4 m_H$ can be measured (among others) from the lattice operator \cite{orblat,MaurizioPaper}
\be
H = {\rm tr} (h^2)\, ,
\ee
since it has the right quantum numbers for a $CP$-even spin 0 scalar.
Out of these two measurements one can form the dimensionless physical ratio $\rho_{HZ}=m_Z/m_H$ that
takes the approximate value $\rho_{HZ}\simeq 1.38$ in the Standard Model and will be important in the following.

As advertised, we observe a pure quantum, bosonic version of the Higgs mechanism, a type
that seems to be rather unusual. Even though this is by itself interesting enough,
a crucial question for us is if this mechanism from the point of view of a four dimensional boundary observer
can look like the SM Higgs mechanism.
An encouraging fact is that the system in the Higgs phase near the bulk first order phase transition {\it simultaneously} reduces
dimensionally to 4d and $\rho_{HZ}$ assumes approximately the experimentally observed value \cite{MaurizioPaper}.
As a consequence, one expects that there is some boundary effective theory (lattice or continuum) with a finite cut-off
that can describe correctly the quantum behaviour of the system. 
It is not hard to guess some basic properties of this effective action.
It must have in its spectrum a $U(1)$ gauge field coupled to a complex scalar (i.e. the Abelian-Higgs model, but this we already knew)
but also at least an additional $Z'$ and a $H'$ state. This is the minimum field content because it has been observed on the lattice \cite{MaurizioProc}.
It could have in principle in addition $Z''$, a $H''$ etc. states, it could have in other words a KK-like tower for each state.
We prefer to call these towers of excited states "KK-like" because non-perturbatively they could have quite different properties from
the classical KK towers that we know. It is now almost trivial to construct the kinetic part of a 4d Lagrangean 
that describes these states. The harder part is to write down the proper scalar potential. This is hard because 
apart from being an effective potential stemming from a non-perturbative mechanism, it must
contain the right information related to the extra dimensional origin of the system,
which means that the couplings in the scalar potential will be necessarily non-generic. 
We can make some progress in this respect by exploiting the fact that an effective potential for the 
lattice Higgs field in \eq{Higgs} can be derived from the gauge invariant expression \cite{symmetries}
\bea\label{Leff}
&& {\cal L}_{\rm eff}^{\rm Higgs} \simeq c_1 \sum_k \tr({\cal Z}_k{\cal Z}_k) + c_2 \sum_k \tr({\cal Z}_k)\tr({\cal Z}_k) 
+ c_3 \sum_{k,l} \tr({\cal Z}_k{\cal Z}_k {\cal Z}_l{\cal Z}_l)\nonumber\\
&+& c_4 \sum_k \tr({\cal Z}_k{\cal Z}_k) \sum_l \tr({\cal Z}_l)\tr({\cal Z}_l) +\cdots
\eea
with $c_1, c_2, c_3, c_4,\cdots$ coefficients undetermined by just the symmetries. 
Expanding the ${\cal Z}_k$ operator above in small lattice spacing and projecting on zero momentum we obtain
\be\label{Veff}
 {\cal L}_{\rm eff}^{\rm Higgs} \supset V_{\rm eff} \sim - \m^2\, {\rm tr} (h^2) + \l \, {\rm tr} (h^4) + \cdots
\ee
where we have defined $\m^2 = 3c_1$ and $\l=9c_3$. This looks like a perfect effective Standard Model Higgs potential
except that $\m^2$ and $\l$ and so also their relative sign are undetermined.
In other words, as far as \eq{Leff} is concerned, we are in the same situation as in the Standard Model where the relative sign that
gives spontaneous symmetry breaking must be input by hand.
This is not so surprising because to construct \eq{Leff} we have used only symmetries.
But we already know from the Monte Carlo analysis of the 5d system (as we discussed above) that the relative minus sign must be the correct one. 
In this sense, the role that experiment plays in the SM, here it is played by the non-perturbative Monte Carlo analysis.

To recap, our guess \eq{Leff} that contains \eq{Veff} is based on symmetries, and the 
relative minus sign necessary to describe a Higgs mechanism is hiding behind the non-perturbative 5d 
dynamics of which we do not have yet an appropriate effective 4d description. 

\subsection{Lines of Constant Physics in the Higgs phase}

A technique to expose in a compact way the quantum behaviour of a system is to
compute its "Lines of Constant Physics" (or LCPs for short). 
Recall that on the lattice the phase diagram is defined in terms of the dimensionless bare couplings.
In our case those are $\b_4$, $\b_5$ and $N_5$ for an effectively infinite four-dimensional volume.
An LCP is then defined as the line on the phase diagram along which at least some of the physical quantities
are kept fixed.\footnote{As a side remark, notice the difference from the usual continuum RG 
statement where the bare couplings after renormalization disappear from the theory and one has an
effective action with renormalized quantities depending on the energy scale.} 
An immediate question is:
how many physical quantities can we keep fixed and how much tuning of the bare parameters is needed in order to do so?
This is a deceivingly simple question to state. 
First of all, since we have 3 dimensionless parameters we can always tune them so that two physical quantities are kept fixed
along a line on the phase diagram. Then, a third, a fourth, a fifth, etc. physical quantity will generally vary from point to
point along the line. Unless if it turns out that some of these are automatically also constant along the line.
This could happen but it is not guaranteed. 

\begin{figure*}[!ht]
\centerline{\includegraphics[width=100mm]{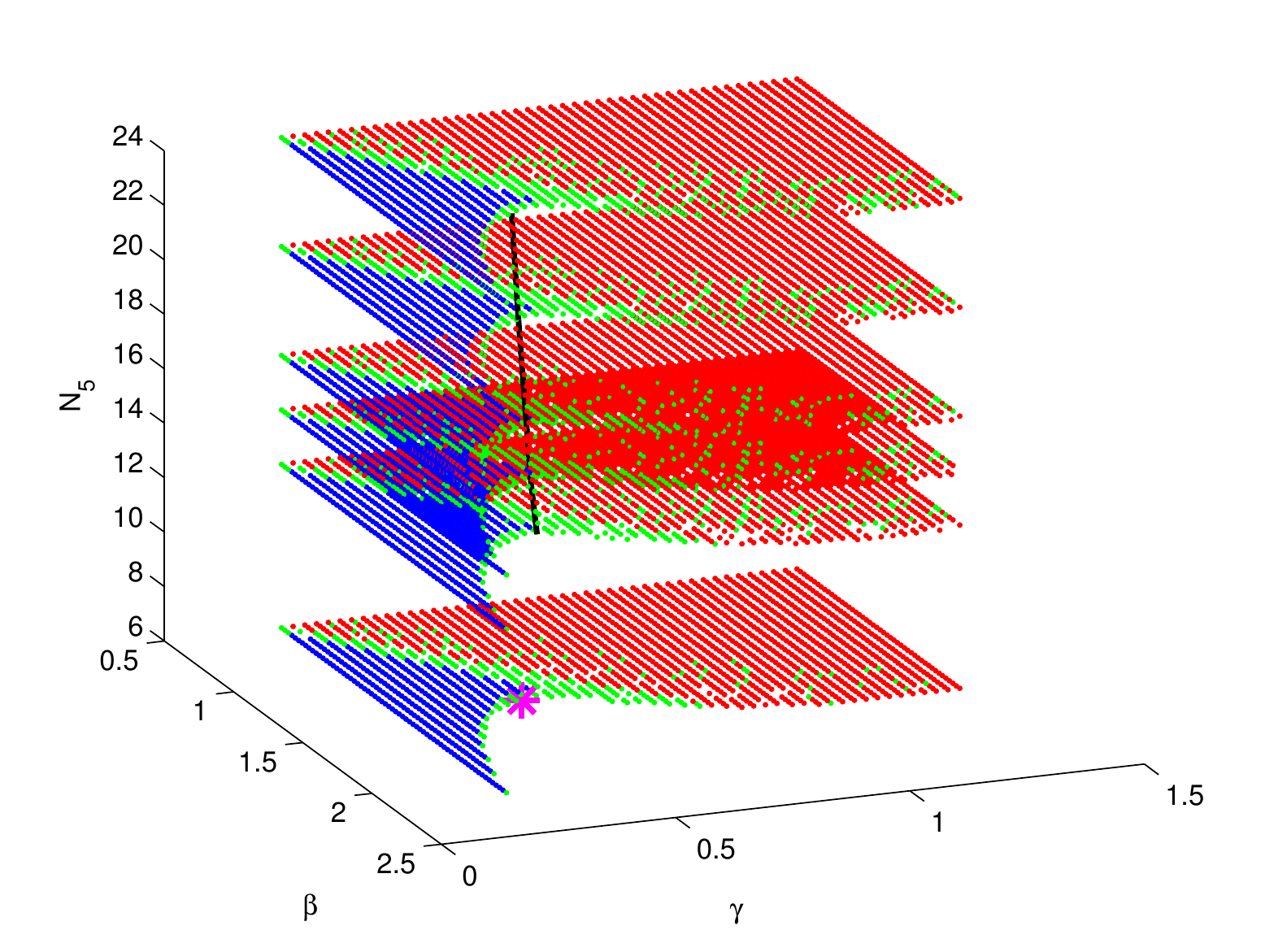}}
\caption{\small{An LCP with $\rho_{HZ}=m_H/m_Z=1.38$ and $m_H R=0.61$, according to a mean-field analysis. It approaches the phase transition for $N_5\to\infty$.
The horizontal axes are $\b_4=\b/\g$ and $\b_5=\b\g$. The quantity $a_4 m_{Z'}$ varies along the LCP.
\vspace{0cm}}}\label{LCP_MF}
\end{figure*}

For example, a Mean-Field (MF) computation \cite{Kyoko} of an LCP on the orbifold lattice, repeated here in Fig. \ref{LCP_MF},
shows that the two quantities $\rho_{HZ}$ and $m_H R$ can be kept fixed but a third quantity such as $a_4 m_{Z'}$ varies along it.
Note that the LCP gets closer to the phase transition as $N_5$ increases.
Now, according to the leading order MF analysis the bulk phase transition is of second order (recall: this is not right, the Monte Carlo analysis shows that the real order is first!).
This allows to make a prediction for $m_{Z'}$ via extrapolation to the continuum limit $N_5\to \infty$, obtaining for a Higgs mass of $125$ GeV, a
$Z'$ mass of $m_{Z'}=989$ GeV.
Of course, we could have chosen other physical observables to keep constant along an LCP and actually the ideal would be to keep
\be\label{LCP}
\rho_{HZ} \simeq 1.38, \hskip 1cm \l \simeq 0.12
\ee
or something similar. The reason that we did not keep $\l$ fixed in the MF example is that
we do not have yet a proper $\l$ lattice operator. 
Nonetheless, let us imagine that we have such an operator in our hands and then it is straightforward to construct the LCP in \eq{LCP}.
Then, the rest of the observables such as the measure of the size of the extra dimension $m_H R$ and 
the excited state ratios $m_{Z'}/m_Z$, $m_{H'}/m_Z$, $\cdots$ will generally vary along the LCP, unless they automatically turn out to be also constant.
And now we have to take a crucial (and very hard) decision:
where on the LCP as it approaches the phase transition is our "physical theory" located?
If the phase transition were of second order the decision would be rather natural, we would
just go all the way until the continuum limit (where the lattice spacing vanishes or in the effective action language the 
cut-off goes to infinity), exactly as we did in our MF example. 
But since the phase transition is really of first order, the lattice spacing may be decreasing along an LCP as the phase
transition is approached but it stops at a non-zero, minimum value when the phase transition is reached. 
The fact that it reduces its value as it approaches the phase transition can be checked in the MF model numerically by 
measuring for example the dimensionless Higgs mass $a_4m_H$ along an LCP closer and closer to the phase transition and
observe that it decreases. 
Since $m_H$ is physical and does not change along an LCP, it must be that $a_4$ is decreasing. 
Based on these numerical facts, we make the following assumption:

{\it {\bf Assumption 1}: The LCP approaches the phase transition with monotonically 
decreasing four-dimensional lattice spacing $a_4$ and terminates on the phase transition.}

So each point along the LCP defines an effective theory with a finite cut-off, whose value is determined by
the value of $a_4$ at that point.
A reasonable guess is that the physical theory sits at a point where the obervables are the smallest in units of the lattice spacing.
This means that the effective theory has a cut-off such that $\Lambda >> m_H, m_Z,\cdots $.
Based on this argument we state our second assumption:

{\it {\bf Assumption 2}: The physical system sits near the phase transition, at or very near the terminal point of the LCP.}

A remaining ambiguity stems from the observation that according to our assumptions,
each point on the phase transition can be considered as the end point of some LCP. 
Moreover, it is not hard to see that the LCP in \eq{LCP} is not unique.
There are possibly more than one LCPs on the phase diagram defined by \eq{LCP},
each with a different value of, say $m_H$. Which one has $m_H\simeq 125$ GeV?
If we find such a point then the LCP that emanates from there and extends into the Higgs phase according to \eq{LCP} is our LCP. 
Then it would be a matter of a few technical steps to find the proper continuum effective action. 
In other words, picking a specific point on the phase transition, say the one that corresponds to 
$m_H=125$ GeV, fixes completely the symmetries, the spectrum and the signs in the effective action as well as the quantum evolution of
the observables.
In fact, there is no possibility other than the boundary effective action
to be the Abelian-Higgs model (plus heavy KK-like states in case it is a continuum effective action) with one important twist: somewhere in the UV the quantum evolution
may have to be modified so that it is properly constrained by the model's higher dimensional nature. 
Precisely this UV regime is hard to reproduce in a 4d effective action because it is where its couplings turn non-perturbative.

The following picture then emerges: in the Higgs phase and far from the phase transition the system is five-dimensional and
perturbative non-renormalizability blurs the picture. All physical quantities along an LCP are of the same magnitude as the cut-off
(meaning $a_4 m_H, a_4m_Z, \cdots \sim O(1)$) and it makes no sense to look for a 4d effective action.
Following an LCP in one direction, that of the lattice spacing going to zero, takes us to
the perturbative point where the 5d coupling goes to zero. This is the point where essentially all 5d model building
takes place and where the CW potential and the continuum Feynman diagram computations have been carried out.
This is also the place where we loose the Higgs mechanism in the pure gauge model and we have to resort to the Hosotani mechanism
(i.e. add fermions etc).
That SSB is lost can be seen also from the fact that the stick symmetry disappears as the lattice spacing goes to zero.
Following an LCP in the opposite direction takes us closer to the phase transition, see Fig. \ref{prx}.
The lattice spacing is again decreasing but never goes to zero as the phase transition is of first order.
In the vicinity of the phase transition the system reduces dimensionally and there is a boundary effective theory
that is basically the 4d Abelian-Higgs model (possibly with some additional excited states) whose low energy
behaviour can be described by usual continuum perturbation theory. This corresponds to the segment $PN$ on the figure.
The UV cut-off however is of the order
of a TeV or so and near the phase transition there should be a regime (see below) where non-perturbative effects must be taken into account,
as the 5d nature of the system affects crucially quantum evolution. 
It is in this regime, segment $NM$ on the figure, where we have to somehow zoom in since by assumption, our physical point is $M$.

\subsection{The other side of the phase transition: the Hybrid phase}

\begin{figure*}[!ht]
\begin{minipage}[ht]{0.5\textwidth}
\begin{flushleft}
\centerline{\includegraphics[width=80mm]{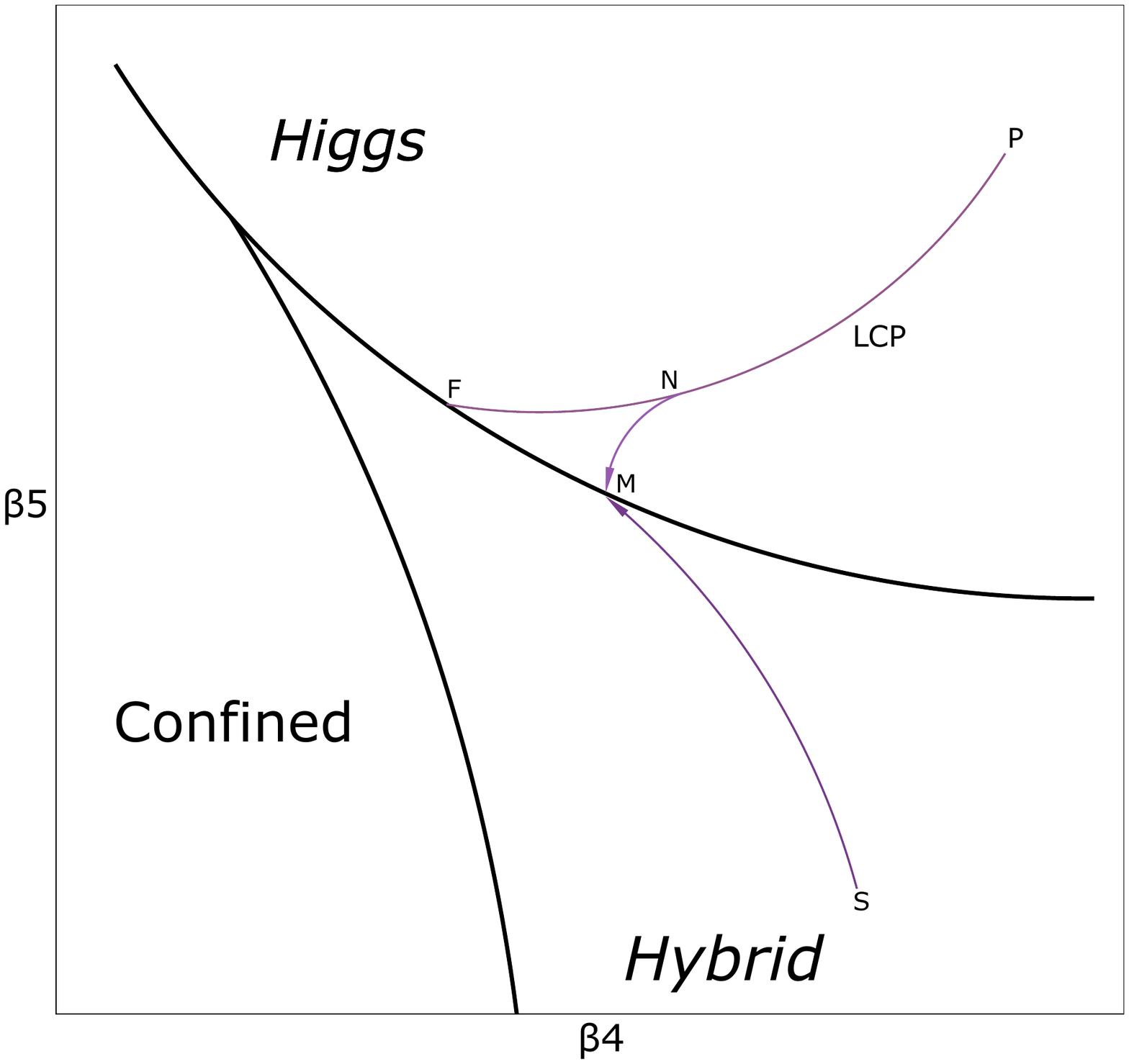}}
\caption{\small{Qualitative Lines of Constant Physics with their $N_5$ dependence projected on the $\b_4$-$\b_5$ plane.
\vspace{0cm}}}\label{prx}
\end{flushleft}
\end{minipage}
\hspace{0.3cm}
\begin{minipage}[ht]{0.5\textwidth}
\begin{flushleft}
\centerline{\includegraphics[width=80mm]{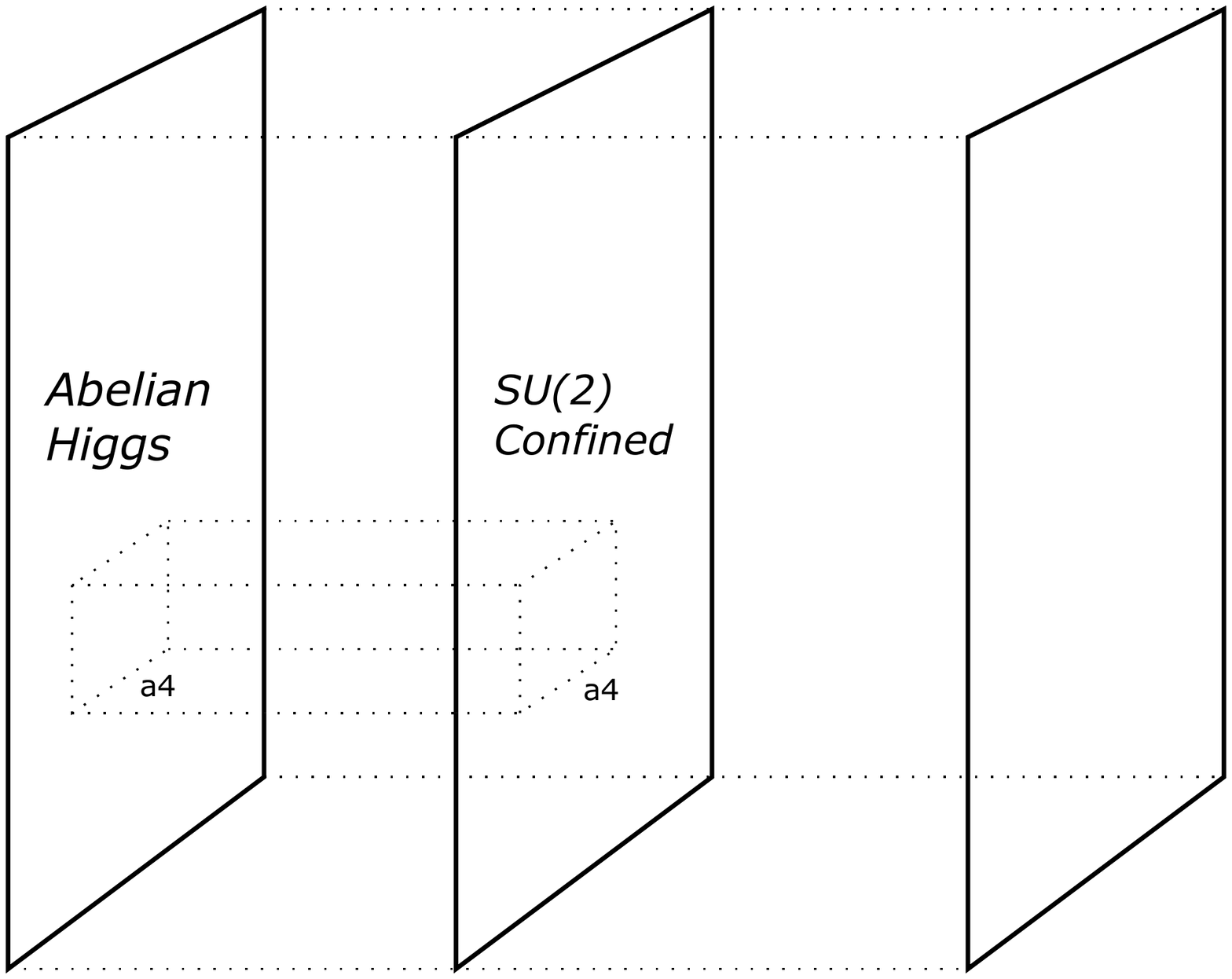}}
\caption{\small{
The orbifold lattice on the phase transition between the Higgs and Confined phases.
\vspace{0.5cm}}}\label{pru}
\end{flushleft}
\end{minipage}
\end{figure*}

On the other side of the phase transition that separates the Higgs from the Hybrid phase the system is in a peculiar state
where the boundary is in a Coulomb phase while the bulk is in a Confined phase \cite{MaurizioPaper}.
This is the reason why we called it Hybrid phase in the first place.
Let us imagine that we sit at a point on the phase transition. At that point, the lattice spacing $a_4$ has a definite value.\footnote{An annoying 
issue that we have to take into account in a careful treatment is that we are sitting on a first order phase transition where 
typically all quantities experience a jump and therefore $a_4$ could be discontinuous.
The region of discontinuity can be probed by measuring expectation values of plaquettes
starting first from a "cold" and then from a "hot" start. The hysteresis observed defines the 
domain of "coexistence" where the system is transforming itself from one phase to the other.
It is useful to think of this state as the water-ice transition that also has such a coexistence state.
For simplicity, we will assume for now that the lattice spacing remains approximately constant while crossing the
coexistence phase.}
We also recall that we are in a regime of the phase diagram where four dimensional slices fluctuate 
approximately independently.
But now we arrive at a surprising conclusion. The point on the line of phase transitions where we are sitting
can be thought of as a point from where two LCP's emanate, one into the Higgs phase (as described above) and one into the Hybrid phase.
The former LCP describes the quantum evolution of the boundary slice and the latter that of one of the bulk slices.
On the phase transition we then have these two theories defined by the same cut-off! See Fig. \ref{pru}.
This is an extremely strong constraint on the dynamics of the system, in particular on the dynamics of the boundary,
in which we are interested.
The most dramatic consequence is that the maximum possible cut-off of the Abelian-Higgs model 
(plus excited states) of the boundary theory is determined by the cut-off of the corresponding confining $SU(2)$ bulk slice.
Because the cut-off of the boundary is tied to the "$\Lambda_{QCD}$" of the 
bulk $SU(2)$ slice, the boundary cut-off can not take very large values and this is consistent with the already
mentioned numerical findings.

\subsection{Towards constructing an Effective Action, }

Next, we speculate on how a more complete picture of NPGHU could look like. This is closely related to being able to write
down an effective action. The effective action could be purely dynamical
or topological or a combination of both. As of today, we are sure only of its dynamical nature.

To probe the dynamics, let us consider two general LCPs, one in the Higgs phase and one in the Hybrid phase, both ending on the same point on the phase transition,
say point M in Fig. \ref{prx}. Actually in the Hybrid phase it does not even need to be an LCP, it could be a usual 
Renormalization Group trajectory.
According to our two Assumptions, as one approaches the phase transition along an LCP in the Higgs phase, flows towards
the UV (segment PN on Fig. \ref{prx}) thus making the Abelian-Higgs system strongly coupled, while approaching it from the side of the Hybrid phase,
one flows towards the UV as well (segment SM on Fig. \ref{prx}), where the bulk $SU(2)$ slice becomes asymptotically free.
Let us assume for the moment that the phase transition is of second order where a continuum limit can be taken.
In this case the asymptotically free running in the Hybrid phase is uninterrupted till the continuum limit
which means that the coupling depends logarithmically on the lattice spacing, for instance at 1-loop according to
\be
a_4 = r_0 e^{-\frac{6\pi^2}{22} \b_4}\, ,
\ee
with $r_0$ the Sommer scale.
This implies that the running on the Higgs phase side must be perturbative all the way until the phase transition
(unless miraculous cancellations leave behind only a logarithmic cut-off dependence).
Consider now the real situation where the phase transition is of first order.
Then, the asymptotically free running in the Hybrid phase (from point S towards point M on Fig. \ref{prx})
is still valid everywhere, except in the vicinity of the phase transition. There, the existence of a finite
cut-off starts being felt by the bulk $SU(2)$ slice and at some point, deviations from the simple 
logarithmic running start developing, for example in the form
\be
a_4 = r_0 \left[e^{-\frac{6\pi^2}{22} \b_4}+ f(\b_4,N_5) \right]\, ,
\ee
where we have assumed that in the Hybrid phase $\b_5$ effects are negligible (as we would expect) and that 
the corrections imposed by presence of the first order phase transition are collected in an additive function $f$.
This means that beyond this point (point N on Fig. \ref{prx}) the running on the Higgs phase side can not be perturbative anymore.
The naive perturbative segment $NF$ should be replaced by the non-perturbative segment $NM$.

Let us now imagine that we draw an LCP starting from a regime of the phase diagram in the Higgs phase where
\begin{itemize}
\item The coupling $\l$ is perturbative
\item The system is dimensionally reduced
\item $m_H=125$ GeV, $\rho_{HZ}\simeq 1.38$ and $\l\simeq 0.12$
\end{itemize}
According to our previous discussion this must be a point already quite close to the phase transition.
This is not unreasonable because recall that $\b_4$ is rather large and $\b_5$ small which means that
the 4d theory is weakly coupled.
In this regime, and up to the point (N on Fig. \ref{prx}) where perturbation theory breaks down, we can use an effective action
determined by the spectrum and the symmetries. In other words, a perturbative Abelian-Higgs model (plus a $Z'$ and $H'$
and perhaps also a $Z''$ and a $H''$ etc.) in a spontaneously broken phase.
An issue here of course is the gauge dependence of the LCP. Assuming that this is under control \cite{Fotis},
we arrive at the point where the perturbative running is not valid anymore. We have the choice to
switch at this point to the full 5d lattice description or to adjust the running in an appropriate way so that
we imitate in the 4d system the non-perturbative 5d running.
The former can be carried out numerically in the context of a Monte Carlo approach \cite{MaurizioProc} or analytically in the context of 
the Mean-Field expansion \cite{Andreas}. The latter can be done by exploiting the relation of LCPs on both sides of the phase transition
as described above.

The deciding fact of a possible topological nature of the Higgs mechanism we have observed is
which of the global symmetries the system chooses to break spontaneously, non-perturbatively, in each phase.
In the Higgs phase, we have already seen the breaking of the stick symmetry that triggers the spontaneous breaking
of the boundary gauge symmetry by the non-zero expectation value of the $CP$-even order parameter ${\cal Z}_k$ in \eq{Zop}.
For $SU(N)$, there are however also $CP$-odd operators \cite{symmetries} that could take
non-zero expectation values. 
Then, the effective action should reflect the broken $CP$ symmetry.
This means that corresponding topologically non-trivial terms could develop in the quantum theory.
It is amusing to notice that in this case the bosonic theory knows something about the
fermions that could be coupled to it, as well as of the way chiral anomaly cancellation would be realised.

{\bf Acknowledgements}

We would like to thank F. Knechtli for useful comments and A. Chatziagapiou and F. Koutroulis for discussions.
We also thank the organizers of the 2016 Corfu Summer School and Workshop for the invitation.

\end{document}